\documentclass[conference]{IEEEtran}
\usepackage{graphicx} 
\IEEEoverridecommandlockouts
\usepackage{cite}
\usepackage{amsmath,amssymb,amsfonts}
\usepackage{algorithmic}
\usepackage{graphicx}
\usepackage{subcaption}
\usepackage{caption}
\usepackage{subfig}
\usepackage{textcomp}
\usepackage{xcolor}
\usepackage{comment}
\def\BibTeX{{\rm B\kern-.05em{\sc i\kern-.025em b}\kern-.08em
    T\kern-.1667em\lower.7ex\hbox{E}\kern-.125emX}}

\newcommand{\bs}[1]{\textcolor{teal}{#1}}
\begin{document}
\title{Goal-Oriented and Semantic Communication in 6G AI-Native Networks: The 6G-GOALS Approach \vspace{-.1cm}\\

\thanks{This work has been supported by the SNS JU project 6G-GOALS under the EU’s Horizon program Grant Agreement No 101139232.}
}

\author{Emilio Calvanese Strinati\IEEEauthorrefmark{1}, Paolo Di Lorenzo\IEEEauthorrefmark{2}, Vincenzo Sciancalepore\IEEEauthorrefmark{3}, Adnan Aijaz,\IEEEauthorrefmark{4}, Marios Kountouris\IEEEauthorrefmark{5}, \\ Deniz G{\"u}nd{\"u}z\IEEEauthorrefmark{6},  Petar Popovski\IEEEauthorrefmark{7}, Mohamed Sana\IEEEauthorrefmark{1}, Photios A. Stavrou \IEEEauthorrefmark{5}, Beatriz Soret\IEEEauthorrefmark{7}, Nicola Cordeschi\IEEEauthorrefmark{2}, \\ Simone Scardapane\IEEEauthorrefmark{2},  Mattia Merluzzi\IEEEauthorrefmark{1}, Lanfranco Zanzi\IEEEauthorrefmark{3}, Mauro Boldi Renato \IEEEauthorrefmark{8}, Tony Quek \IEEEauthorrefmark{9},\\ Nicola di Pietro\IEEEauthorrefmark{10}, Olivier Forceville\IEEEauthorrefmark{11}, Francesca Costanzo\IEEEauthorrefmark{1}, Peizheng Li\IEEEauthorrefmark{4}
\smallskip\\
\IEEEauthorrefmark{1}CEA-Leti, Université Grenoble Alpes, Grenoble, France,
\IEEEauthorrefmark{2}CNIT, Italy\\
\IEEEauthorrefmark{3}NEC Laboratories Europe, Heidelberg, Germany,
\IEEEauthorrefmark{4}Toshiba Europe Ltd., Bristol, United Kingdom\\
\IEEEauthorrefmark{5}Communication Systems Department, EURECOM, Sophia-Antipolis, France\\
\IEEEauthorrefmark{7}Department of Electronic Systems, Aalborg University, Aalborg, Denmark\\
\IEEEauthorrefmark{6}Department of Electrical and Electronic Engineering, Imperial College London, \IEEEauthorrefmark{8} TIM, Italy\\
\IEEEauthorrefmark{9}Singapore University of Technology and Design, Singapore\\
\IEEEauthorrefmark{10} Hewlett Packard Enterprise, Italy, \IEEEauthorrefmark{11} Hewlett Packard Enterprise, France

\vspace{-.3cm}}

\maketitle

\begin{abstract}
Recent advances in AI technologies have notably expanded device intelligence, fostering federation and cooperation among distributed AI agents. These advancements impose new requirements on future 6G mobile network architectures. To meet these demands, it is essential to transcend classical boundaries and integrate communication, computation, control, and intelligence. This paper presents the 6G-GOALS approach to goal-oriented and semantic communications for AI-Native 6G Networks. The proposed approach incorporates semantic, pragmatic, and goal-oriented communication into AI-native technologies, aiming to facilitate information exchange between intelligent agents in a more relevant, effective, and timely manner, thereby optimizing bandwidth, latency, energy, and electromagnetic field (EMF) radiation. The focus is on distilling data to its most relevant form and terse representation, aligning with the source's intent or the destination's objectives and context, or serving a specific goal.
6G-GOALS builds on three fundamental pillars: \emph{i}) AI-enhanced semantic data representation, sensing, compression, and communication, \emph{ii}) foundational AI reasoning and causal semantic data representation, contextual relevance, and value for goal-oriented effectiveness, and \emph{iii}) sustainability enabled by more efficient wireless services. Finally, we illustrate two proof-of-concepts implementing semantic, goal-oriented, and pragmatic communication principles in near-future use cases. Our study covers the project’s vision, methodologies, and potential impact.

\end{abstract}
\smallskip

\section{Introduction}
For decades, communication system design has progressed through incremental cycles aiming to increase the network throughput and coverage, and, in 5G, to provide massive and/or ultra-reliable connectivity. This has been achieved mainly by increasing the bandwidth, or introducing new network elements, such as multiple antennas, small cells, or, more recently, reconfigurable intelligent surfaces (RIS) to opportunistically shape the propagation in wireless environments \cite{CalvaneseRISE-6G2021}. Despite the constant growth in data traffic, these measures have been successful in meeting the increasing demand so far. However, as we are going towards 6G networks, we are entering a new phase in communications geared to natively interconnect Artificial Intelligence (AI) modules \cite{LIU202242,HexaX6G2021} in a sustainable way, i.e., by avoiding the paradox of increasing efficiency while experiencing a much higher increase of data traffic. The upcoming 6G networks are expected to create a network of networks through the convergence of communication, computation, control, and learning principles, supported by efficient interactions and exchange of knowledge among agents with diverse forms of intelligence \cite{calvanese2019,6GLetaief2019}. 

\begin{figure*}
    \centering
\includegraphics[width=0.89\textwidth]{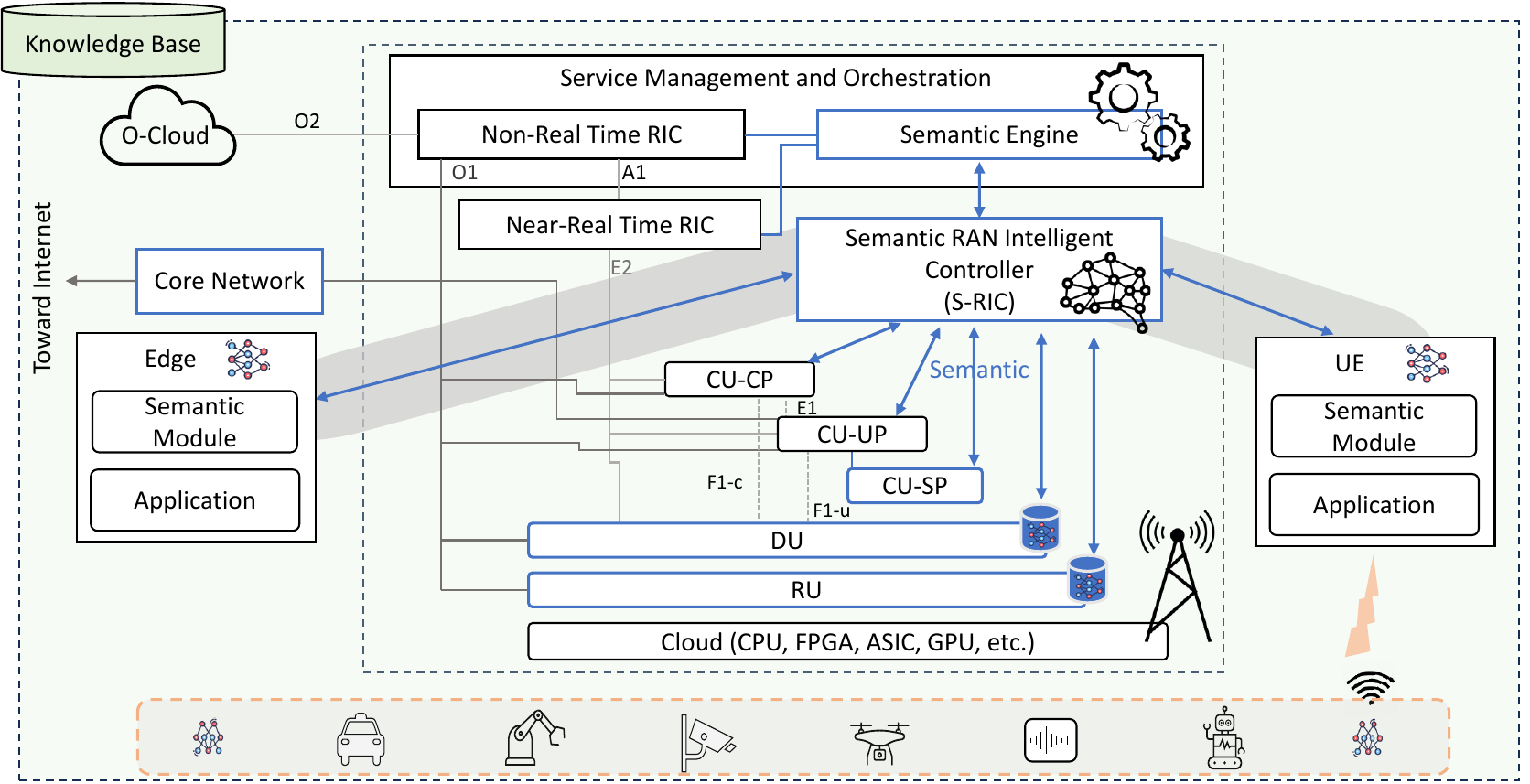}
    \caption{Illustration of the 6G-GOALS system architecture.}
    \label{arch}
\end{figure*}

\textbf{Project vision:} The 6G-GOALS project aims to realize the potential of cutting-edge AI-native architectures along with new semantic and goal-oriented communication paradigms. The project challenges the prevalent method of transmitting data without understanding its relevance or informativeness for AI algorithms, a process described as \textit{content-blind-transmit-without-understanding}. Contrary to the common belief that receivers are always interested in the transmitter's message, this paper emphasizes that information is valuable only when it holds practical, exploitable significance for the receiver. This approach targets the transmission of only the most pertinent information, tailored to the specific learning or inference objectives in machine-oriented communications. 

The conceptual groundwork for this approach was laid over 70 years ago by Shannon and Weaver~\cite{shannon1948mathematical}, who identified three levels of communication: the syntactic (wireless) level, the semantic (meaning) level, and the effectiveness (goal) level.
Recently, a new surge of interest in semantic and goal-oriented communication has emerged, identifying this new paradigm as a keystone for exploring the meaning behind the bits and enabling brain-like cognition and effective task execution among distributed network nodes \cite{CalvaneseGOWSC2021}. Moreover, novel approaches to semantic joint source and channel coding \cite{Deniz2023GOJSSC}, semantic plane in 6G architectures  \cite{popovski2020semantic}, semantic extraction and compression \cite{kountouris2021Semantic,stavrou2023role,SANACemComCCNC22}, goal-oriented system design and optimization \cite{di2023goal}, semantic reasoning \cite{ThomasCCNC2024,xiao2023reasoning}, and semantic communications based on generative AI \cite{barbarossa2023semantic} have been proposed. 


While a theory of semantic and goal-oriented communications is currently absent, the project seeks to lay the groundwork for semantic goal-oriented communication. This includes exploring new methods to represent, sense, compress, and convey semantic information. The objectives are to bridge research efforts into concrete practical implementation and standardizable industrial use in 6G, to benchmark their potential advantages for 6G against classical, semantic-agnostic approaches, and to explore solutions for sustainable and effective coexistence with legacy communication systems (i.e., data-oriented) \cite{MerluzziGO2023}. To this end,
the 6G-GOALS project proposes a novel AI/ML-based architecture built on top of hierarchical O-RAN architecture where the semantic plane, \textit{semantic RAN intelligent controllers} (S-RIC) and \textit{semantic reasoning engines} units are introduced.
Moreover, we investigate the foundations of semantic-based distributed reasoning and goal-oriented actuation in time-varying data scenarios. First, we assume that the interaction between agents is based on an agreed common semantic setting, thus assuming the use of a common ontology language to represent knowledge and/or compatible semantic extraction (at the transmitter) and interpretation (at the receiver) logic/models. Here we incorporate causal inference in representation learning to allow efficient semantic representation of unstructured data and also to robustify learned models with respect to possible distribution shifts. 
Second, going beyond the state of the art and following the initial results in \cite{HuttebrauckerCemComCCNC24,SANAGlobecom23}, we investigate the impact and potential countermeasures when distinct semantic settings lead to communication errors at the semantic and effectiveness levels. 
These errors may induce interpretation errors and result in defective AI cooperation strategies, potentially affecting the communication's effectiveness. 
The proposed framework enables the exploration of novel resource allocation strategies for effective semantic-oriented communications, aiming for substantial energy and spectrum efficiency, reduced EMF exposure, and interoperability of massive machine-type communications, while ensuring backward compatibility with semantic agnostic agents.

\section{6G-GOALS System Architecture} 

The system architecture for 6G-GOALS is built on the O-RAN architecture \cite{o2021ran}, which is a key enabler of innovation in 6G \cite{O-RAN_6G_nGRG} and offers several benefits in terms of underpinning semantic and goal-oriented communications \cite{li2023open_arxiv}. The essential design elements for the 6G-GOALS system architecture are outlined as follows:
\begin{itemize}
\item Establishing an AI-native 6G system specifically tailored for semantic and goal-oriented communications.
\item Offering a flexible design that includes intelligent network functions. These functions will be responsible for semantic data extraction, information representation, and managing and optimizing resources.
\item Implementing a semantic plane that enhances both the user plane and the control plane. This implementation aims to facilitate the delivery of semantic services, enable knowledge-driven reasoning, and improve both user experience and overall system efficiency.
\item Creating an intelligent and adaptable Radio Access Network (RAN) to effectively handle semantic and goal-oriented communication on a large scale.
\item Ensuring that the new semantic-aware system components can operate in conjunction with existing legacy system components.
\end{itemize}

The envisaged 6G-GOALS system architecture is illustrated in Fig. \ref{arch}. It leverages the programmability, the disaggregation, and the openness, and more importantly, the  RAN intelligent controllers (RICs), offered by the O-RAN architecture for achieving the aforementioned design requirements. The overall architecture considers a disaggregated RAN, i.e., the base station is split into the radio unit (RU), the distributed unit (DU), and the centralized unit (CU). 

\subsection*{6G-GOALS Architectural Blocks}

The key components of the system architecture in Fig. \ref{arch} are described in the sequel.

\textit{Semantic Engine} -- The semantic engine is responsible for the effective and efficient delivery of semantic-oriented services through the orchestration of semantic information resource processing, lifecycle management of semantic models, and user experience management. In terms of hierarchy, the semantic engine is positioned at the service management and orchestration (SMO) layer of the overall system.  

\textit{Semantic RIC} --  The Semantic RIC provides a programmable and extensible platform for the deployment of semantic-oriented applications (semantic xApps). It is connected to the control and user planes of the CU as well as to the DU. This connectivity can be realized via existing O-RAN-specified interfaces or new interfaces that are being investigated in the project. The semantic RIC  supports different service time requirements, ranging from non-real-time to near-real-time and real-time.

\textit{RAN Semantic Plane} -- The semantic plane in the RAN spans the CU, the DU, and the RU. The semantic plane in the CU, i.e., the CU-SP, takes instructions from the semantic RIC and communicates with the CU user and control planes, i.e., CU-UP and CU-CP, respectively, for effective task coordination. Further, it handles the semantic information flow with the DU and the RU. The DU and the RU are semantic-aware and these components can co-exist with legacy components. The semantic awareness in the DU and the RU implies empowering these components with a range of functionalities such as semantic information extraction and encoding and JSCC. Further, the semantic-aware RU and DU should be coordinated by a set of well-defined semantic model management/update policies controlled by the semantic engine and the semantic RIC. 

\textit{Application Plane} -- The application plane is orthogonal to the semantic plane and provides necessary interfaces for the semantic applications across edge devices or user equipment. 


\textit{UE and Edge Components} -- 
The User Equipment (UE) and edge devices will be significantly upgraded with advanced computational and learning abilities. This enhancement is crucial for the initial phase of semantic information extraction, where they will be tasked with processing and interpreting raw data from various sources such as images, videos, and sensor inputs. This will be possible thanks to a sophisticated semantic module capable of not only extracting meaningful information from these raw data formats but also understanding and contextualizing them to provide a more intelligent and efficient data processing workflow.


\textit{Knowledge Base} -- The knowledge base is a cornerstone of this architecture, seamlessly involved in virtually all aspects of semantic processing modules. Its involvement is crucial across all facets of semantic model development, from training to validation. The presence of a coherent and unified knowledge base is vital for efficiently decoding and interpreting semantic information, providing support for these processes.

\textit{Enhanced Core Network} -- The core network can further support the semantic-empowered functionalities of 6G-GOALS for the RAN and the management, orchestration, and application domains. For instance, the network data analytics function (NWDAF) can collect performance metrics for AI-based optimization in the semantic engine and the semantic RIC. The network exposure function (NEF) can expose network control and management features to the semantic engine and the semantic RIC, in terms of policy control, QoS management, traffic re-routing, etc. 

It is emphasized that the architecture supports generative AI. For example, generative AI fine-tuned for vertical applications (with a specific knowledge base) can be deployed in the semantic engine. The generative models after pruning have the potential to be deployed at the RU/DU level. 


\section{Research on Innovative Technology Enablers} 
The 6G-GOALS framework aims to go beyond the established limits of the current sense-compute-connect-control models and transition toward semantic communication-based AI architectures, protocols, and services. The envisioned pillars are described in the sequel and can be summarized as (i) AI-empowered semantic data representation, sensing, compression, and communication; (ii) timing-aware semantic communication for distributed reasoning and actuation; (iii) 6G sustainability via semantic-empowered RAN.

\subsection{AI-empowered semantic data representation, sensing, compression, and communication}
The first pillar of 6G-GOALS will be the development of the theoretical and algorithmic foundations of semantic and goal-oriented communications, encompassing several key aspects such as data acquisition, extraction, representation, compression, and caching to minimize the amount of data to be transmitted to optimize desired performance metrics under complexity and physical constraints. The methodologies are detailed in the sequel.
\subsubsection{Mathematical Definitions and Fundamental Limits}
6G-GOALS focuses on developing information-theoretic fundamentals for networked systems, targeting semantic characteristics and information representation for improved cost and performance. It involves analyzing network trade-offs, studying semantic metrics, and assessing the value of information \cite{kountouris2021Semantic,stavrou2023role}. The output covers multiuser semantic communication, including sampling and compression, and explores efficient computational methods using both convex and non-convex optimization. Additionally, it examines cost-performance trade-offs in semantic communication and strategies for energy-saving and complexity reduction using generative AI models and theoretical approaches like Kolmogorov complexity. The project also investigates topological models to balance data representation richness with semantic information extraction.
\subsubsection{Semantic Data Acquisition, Representation, and Compression} 6G-GOALS aims to design and optimize algorithms for semantic and goal-oriented data acquisition, representation, and compression, targeting relevance for specific tasks while considering energy, accuracy, and latency constraints. Objectives include developing compact semantic data representations for effective communication, using manifold learning and topological neural networks to uncover key information properties \cite{battiloro2023latent}, and employing theoretical methods for optimal representations. It also involves designing AI algorithms for rate-limited channels and addressing the rate-distortion-perception trade-off, along with delivering large neural networks, like Bayesian networks, with training and fine-tuning methods considering compression rates. Additionally, the task focuses on semantic sensing \cite{kountouris2021Semantic}, utilizing active learning, joint sampling and transmission techniques, and sparsity-inducing attention mechanisms for relevance detection in data.

\subsubsection{Semantic Source and Channel Coding Schemes}
The focus of this activity is on Joint Source-Channel Coding (JSCC), particularly using deep neural networks (DeepJSCC) \cite{Bourtsoulatze:TGCN:19, xu2023deep}, to extract, process, and transmit data content like images and videos. The goal is twofold: to advance DeepJSCC beyond current convolutional models in terms of complexity, training, and generalizability, and to integrate these advancements into practical solutions for next-generation mobile networks. This involves: \emph{i}) creating modular JSCC architectures compatible with existing standards and tailored for industry-specific applications; \emph{ii}) developing transformer-aided DeepJSCC for efficient media transmission; \emph{iii}) designing DeepJSCC for semantic communications using topological DNNs; \emph{iv}) devising adaptable JSCC schemes for various signal sources. Additionally, the project will exploit wireless delivery of DNN models, a critical aspect of JSCC, providing also a proof of concept implementation (cf. Sec. IV.A). 

\subsection{Timing-aware semantic communication for distributed reasoning and actuation} 
The second pillar of 6G-GOALS entails distributed reasoning and actuation in timing-varying data scenarios. The main outputs of this research area are AI/ML-aided semantic communication techniques and reasoning algorithms that respect timing constraints when executing distributed coordination and control to satisfy the KPIs defined by 6G-GOALS.

\subsubsection{Foundations of Reasoning through Semantic-Empowered Communications}
In semantic communications, agents typically use compatible languages and models, with errors usually due to lossy compression or incomplete knowledge. However, using different languages and models can lead to semantic noise and performance issues~\cite{HuttebrauckerCemComCCNC24,SANAGlobecom23}. 6G-GOALS will tackle these challenges using domain adaptation theory to address language and model mismatches between senders and receivers. The methodology will be based on optimal transport theory for effective semantic recovery, linking it to domain adaptation and rate-distortion theory, and on JSCC schemes that are robust against model mismatches.

\subsubsection{Semantic Communications under Timing Constraints} 6G-GOALS aims to develop new methodologies that focus on the relationship between context, relevance, and value of information, utilizing advanced time-aware metrics ~\cite{Deniz2023GOJSSC, gunduz2023timely, oranus}. The methodology includes: \emph{i}) Techniques to represent information relevance and capture time-dependent aspects; \emph{ii}) An AI/ML framework to optimize data transmission and decision-making within specific timing constraints; \emph{iii}) Dynamic information exchange strategies using AI/ML for data relevance and timing-based communication; \emph{iv}) Stochastic optimization and reinforcement learning methods for effective policy development under timing constraints; \emph{v}) Context-specific coding schemes and prioritized communications.

\subsubsection{Causal Semantic Representations and Pragmatic Communications}

Traditional ML-based communication systems, which often ignore causality in favor of statistical models, struggle with real-world data distribution shifts. 6G-GOALS aims to move towards causal models for a better understanding of interventions and planning in semantic communications. It involves learning causal representations from time-varying data, focusing on extracting causal structures from unstructured data, and employing multi-view learning for causal semantic data representation. The project will use advanced techniques from Markov decision theory, reinforcement learning, and optimal transport theory to address sequential optimization problems with causal objectives \cite{stavrou:2022}. This includes developing scalable causal estimation and low-delay compression techniques for semantic message reconstruction and exploring energy-efficient AI/ML models. The aim is to create interventions in 6G-GOALS systems as sequences of communication messages and actions, extending causal reinforcement learning to a distributed setup for improved pragmatic communication and action strategies.

\subsection{6G sustainability via semantic-empowered RAN}
The third pillar of 6G-GOALS aims at demonstrating the benefits of semantic communication, bridging architectural, theoretical, and algorithmic solutions with operationally adaptive intelligent RAN mechanisms. To this end, consolidated and novel semantic-performance metrics will be instrumental in devising mechanisms that optimize the new emerging trade-offs among several KPIs, such as a) spectrum efficiency, b) energy efficiency, c) reliability, d) latency, e) performance of end-to-end communication, learning and control.

\subsubsection{Semantic control strategies}
6G-GOALS aims to develop efficient L1/L2 protocols and control signaling for ML and semantic communication in classical and ORAN architectures. It involves creating a semantic radio intelligent controller (S-RIC) to enhance real-time control and network efficiency. The approach includes implicit transmission of control information using AI, reducing overhead and improving system resilience. Key efforts include designing L1/L2 protocols for two-way communication, focusing on energy and overhead efficiency, and integrating semantic control into the O-RAN architecture. Additionally, we explore new semantic control metrics for error handling and message reformulation, ensuring compatibility with existing vendor-specific systems.

\subsubsection{Connect-Compute Network Resource Management for Semantic Communications}
6G-GOALS aims to design and optimize network resource allocation for semantic-oriented communications, focusing on energy and spectrum efficiency. It involves balancing performance between semantic and traditional communication systems through innovative spectrum management and AI-driven resource management. Key areas include developing new metrics for spectrum sharing, prioritizing goal-effectiveness and semantics, and using AI for energy-efficient management of network resources. The approach includes minimizing data transmission while maintaining communication goals. Additionally, Over-the-Air Computing (OAC) methods will be explored to compute arbitrary functions from multiple transmissions efficiently, moving beyond traditional decoding methods.

\begin{figure}[t]
    \centering
\includegraphics[width=\columnwidth]{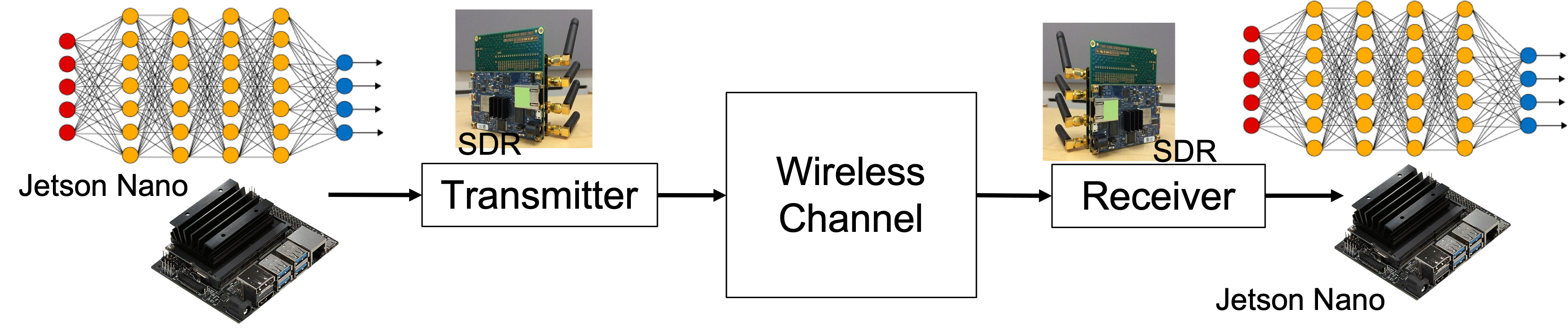}
    \caption{Experimental setting of PoC on real-time semantic communication.}
    \label{fig:poc1}
\end{figure}

\section{Proofs of Concept} 
Realistic use cases will substantiate the attainable performance of semantic and goal-oriented communications while pursuing different objectives and comparing them with conventional communication schemes. This will be performed and hereafter described using two different PoCs involving: \emph{i}) hardware implementation of semantic-aware communication; and \emph{ii}) semantic-aware collaborative robots.

\subsection{Hardware implementation of semantic communication}
This PoC focuses on implementing a real-time semantic communication system over a wireless channel with bandwidth and power constraints, focusing on a point-to-point communication link. In particular, we will focus on the transmission of a model for a learning task, e.g., the delivery of a neural network model over a wireless channel \cite{Jankowski:ISIT:22}. The transmitter has the goal of enabling the receiver terminal to perform inference on new data samples by exploiting a semantic communication paradigm. The hardware implementation will equip terminals with JetsonNano processors for training and inference of large AI/ML models, as depicted in Fig.\ref{fig:poc1}. Additionally, software-defined radio (SDR) units will be utilized to map model parameters directly to channel symbols, providing a flexible and cost-effective implementation tested under various channel constraints and mobility conditions.

\begin{figure}[t]
    \centering
    \includegraphics[width=0.49\textwidth]
    {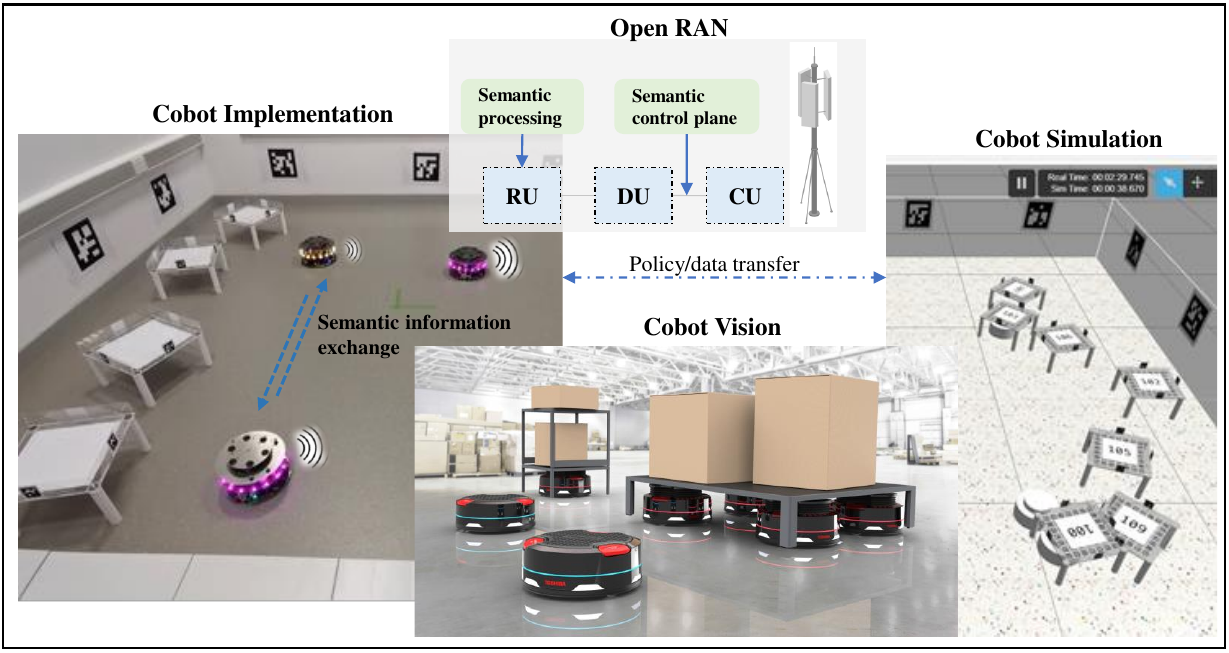}
    \caption{Semantic-aware cooperative robotics PoC.}
    \label{fig:poc2}
\end{figure}

Two scenarios are considered: in one, data samples are wirelessly transmitted to the receiver for model training; in the other, a model is trained at the transmitter terminal and sent to the receiver. Due to bandwidth and latency constraints, samples in the first approach and the model in the second approach are transmitted in a lossy manner, contingent on the channel state. The primary focus is on a joint source-channel coding approach for model transmission, where model parameters directly map to OFDM symbols, and the receiver utilizes noisy model parameters. To enhance model robustness against channel noise, noise injection during training and distillation methods will be employed. 


\subsection{Semantic Communications-enabled Cooperative Robotics}

The principal objective of this PoC is to demonstrate an improved connectivity layer driven by semantic and goal-oriented communication for a cooperative robotics use case, showcasing and measuring the potential advantages of semantic communications in such scenarios. Conducted in a specialized arena with multiple low-cost multi-radio, multi-sensory robotic platforms designed for collaborative operations in warehousing and logistics, the trial will assess the performance of a semantic communications-based approach compared to the conventional method relying on the transmission of raw command/feedback messages. A qualitative scenario example is illustrated in Fig. \ref{fig:poc2}. The connectivity layer for the robotic platforms will be provided by an advanced 3GPP-compliant core network (Rel-17/18) and an advanced 5.5G Open RAN system ~\cite{BEACON_Open_RAN} equipped with semantic awareness and an enhanced RIC platform. Leveraging GPU capabilities, the robot nodes will serve as semantic information processing nodes, encoding or decoding such information in real-time from a heterogeneous set of sensors, including inertial units, LiDAR, camera, etc. Emphasis will be placed on the exchange of sensory data, with a semantic-aware physical layer designed to effectively manage sensor data. A comprehensive semantic knowledge base will be established for managing the semantic codec, and the in-lab demonstration of semantic model training and deployment in edge devices will be integrated into the cobots testbed for large-scale deployment. The disaggregated O-RAN architecture will play a crucial role in realizing semantic awareness enhancements, offering flexibility for the implementation of semantic communications. The testbed will be enriched with a semantic-aware controller for the multi-robot system to efficiently manage semantic models and provide in-time semantic responses. Moreover, the O-RU component will be empowered with near real-time semantic processing capability for robotic sensor data in conjunction with the introduction of the semantic control plane. To effectively support semantic communication, control, and task allocation in cooperative robotics systems all the software will be developed and instantiated as virtualized instances, e.g., microservices, xApps, rApps, etc.
Beyond RAN-centric enhancements. The core network will also undergo improvements to support semantic communications, including the introduction of new quality-of-service classes to facilitate end-to-end semantic communication and the exposure of new application functions for policy-based control of semantic-aware network components.

\section{Conclusions}
This paper presents the vision of the 6G-GOALS project, aiming to establish the theoretical, algorithmic, and practical cornerstones for semantic and goal-oriented communication. This vision concentrates on managing semantic data--its representation, sensing, compression, and communication--while delving into the basics of semantic-aware AI reasoning. This involves comprehending causal semantic data in its time context, emphasizing relevance and value for efficient goal-oriented communication, and exploring the effects of different semantic communication scenarios among cooperative AI agents. Then, to transition research into practical, standardizable industrial applications for 6G, the paper identifies key technical challenges: \emph{i}) creating a new network architecture incorporating a semantic plane and specialized semantic modules; \emph{ii}) determining the essential limits of semantic and goal-oriented communications; \emph{iii}) crafting sustainable, efficient strategies to minimize connect-compute-sense resources while maintaining compatibility with existing data-oriented communication systems; and \emph{iv}) developing experimental benchmarks through two distinct proof-of-concept scenarios.

\bibliographystyle{IEEEtran}
\bibliography{bibliography}

\end{document}